\documentclass[apsprb,longbibliography, floatfix,secnumarabic,amssymb,nobibnotes,nofootinbib,showpacs]{revtex4-2}

\usepackage[T1]{fontenc}
\usepackage[utf8]{inputenc}
\usepackage{amsmath,amssymb,amsfonts}
\usepackage{graphicx}
\usepackage{braket}
\usepackage{physics}
\usepackage{hyperref}
\usepackage{xcolor}
\usepackage{bm}     
\usepackage{placeins}
\usepackage{subcaption}  

\usepackage{grffile}
\usepackage{float}
\usepackage{caption}
\graphicspath{{./}{./figs/}} 
\hypersetup{
    colorlinks = true,
    citecolor  = blue,
    linkcolor  = blue,
    urlcolor   = blue
}

\begin{document}

\title{Controlling correlations of a polaritonic Luttinger liquid by engineered cross-Kerr nonlinearity}
\author{Nabaneet Sharma, Anushree Dey and Bimalendu Deb}
\email{msbd@iacs.res.in}
\affiliation{School of Physical Sciences, Indian Association for the Cultivation of Science, Jadavpur, Kolkata 700032, India.}

\begin{abstract}
We study correlation properties of polaritons at zero temperature in a multiconnected Jaynes--Cummings (MCJC) lattice on a superconducting circuit quantum electrodynamics platform with engineered cross-Kerr nonlinearity that mimics attractive nearest-neighbour interaction. A multi-connected Jaynes--Cummings lattice is a one-dimensional lattice constructed from alternating qubits and resonators with different left and right couplings. The nearest-neighbour interaction or cross-Kerr coupling is implemented dispersively through ladder-type qutrits between each nearest neighboring pair of resonator modes. Projecting onto the lower-polaritonic manifold, we derive an extended two-mode (bipartite) Bose--Hubbard-like model featuring on-site and attractive nearest-neighbor interactions. Employing a continuum bosonization approach, we express the Hamiltonian in terms of symmetric ($+$) and antisymmetric ($-$) collective modes. In the regime where the ($-$) sector acquires a finite gap, one can reduce the system to an effective single-component Luttinger liquid model for the $+$ sector. The cross-Kerr term reduces the compressibility of the ($+$) mode, thereby enhancing the corresponding Luttinger parameter $K_{+}$, resulting in the slower algebraic decay of single-particle correlations, $G(x)\propto|x|^{-1/(4K_{+})}$.
\end{abstract}

\maketitle
\section{Introduction}

Superconducting circuit quantum electrodynamics (cQED) lattices have emerged as an excellent platform for exploring new models of strongly interacting photonic or polaritonic many-body physics with microscopic, in-situ control over couplings, nonlinearities, and dissipation \cite{Hartmann2006NatPhys,Greentree2006NatPhys,Du2024FundRes,puertas2019tunable,carusotto2020photonic}.  In one dimension, this platform is especially valuable because universal low-energy descriptions often apply, enabling direct connections between circuit parameters and long-distance coherence \cite{fitzpatrick2017observation,ma2019dissipatively}. A central goal is to engineer interactions beyond purely on-site Kerr terms so that a lattice can emulate extended Hubbard models and display density ordering, nontrivial insulating regimes, and tunable correlations \cite{roushan2017chiral,owens2018quarter}.

Nonlinear photonics has a long history across atomic and cavity platforms, ranging from electromagnetically induced transparency schemes that yield large Kerr responses \cite{Schmidt1996OL,Imamoglu1997PRL} and conditional phase shifts at the few-photon level \cite{PhysRevLett.75.4710, PhysRevLett.91.093601}.  In superconducting microwave circuits, such nonlinear interactions can be engineered and measured with high precision, including  cross-Kerr effects for propagating fields \cite{PhysRevLett.127.050502,
kounalakis2018tuneable} and single-photon-resolved cross-Kerr interactions suitable for autonomous stabilization of photon-number states \cite{chapple2025balanced}. Cross-Kerr nonlinearities have also been explored in optomechanical settings, providing another route to effective photon–photon interactions \cite{PhysRevA.93.023844,ye2024ultrafast}. More recently, nonperturbative cross-Kerr couplings have enabled fast, high-fidelity quantum nondemolition readout protocols, underscoring the importance of cross-Kerr effects as a resource in cQED.

A particularly appealing lattice architecture is the multiconnected Jaynes--Cummings (MCJC) chain, where qubits mediate connectivity between resonators such that effective kinetic processes arise from light--matter coupling itself rather than explicit photon hopping \cite{Xue2017,schmidt2013circuit,majer2007coupling,PhysRevA.76.042319}.  At (or near) zero qubit--resonator detuning, the natural quasiparticles are Jaynes--Cummings polaritons, which can be systematically mapped to an effective bosonic lattice model \cite{RevModPhys.85.299,noh2017quantum}.  While the intrinsic Jaynes--Cummings nonlinearity generates an on-site interaction, to access highly correlated regime in one dimension  requires engineered interactions between neighboring sites\cite{hafezi2011robust}.  Cross-Kerr density--density couplings of the form $\chi\,n_i n_{i+1}$ represent nearest-neighbor interactions in an extended Bose--Hubbard (EBH) language and can be implemented dispersively in circuit QED architectures \cite{PhysRevLett.111.053601}.

The EBH model in one dimension supports a variety of phases that are absent in the on-site-only Bose--Hubbard problem, including charge-density-wave order and topologically nontrivial insulating behavior such as the bosonic Haldane insulator \cite{Ejima2011PRL}. Related extensions with cavity-mediated long-range interactions have also been discussed in the extended Bose–Hubbard context, including their impact on insulating regimes \cite{SicksRieger2020EPJB}. Additional terms---for example pair-hopping processes---further enrich the phase structure and the competition between coherent transport and interaction-driven ordering \cite{PhysRevA.88.063613}.  Away from commensurate pinning, however, the long-distance behavior of a broad class of one-dimensional interacting systems is described by Tomonaga--Luttinger liquid (LL) theory, in which correlation functions decay algebraically with exponents set by Luttinger parameter $K$ and sound velocity $u$ \cite{Haldane1981JPC,Giamarchi1991PRB,GiamarchiBook2003}, for connecting microscopic lattice models to continuum observables \cite{Imambekov2012RMP,PhysRevResearch.3.023062}.

In this work, we develop a continuum LL description of an MCJC lattice with \emph{attractive} nearest-neighbor interactions.  Starting from a microscopic circuit-QED Hamiltonian with asymmetric couplings and a dispersively generated cross-Kerr interaction, we project the system onto the lower-polariton manifold at zero detuning and derive an effective \emph{bipartite} EBH model featuring on-site repulsion together with an attractive nearest-neighbor term $-\chi\,n_i n_{i+1}$.  Employing bosonization, we formulate the low-energy theory in terms of symmetric $(+)$ and antisymmetric $(-)$ collective modes.  In the regime where the $(-)$ sector acquires a finite gap, one can  integrate out the $(-)$ mode, yielding an effective single-component LL description for the $(+)$ sector.  Within this analytically tractable regime, we show that attractive cross-Kerr interactions reduce the effective compressibility penalty in the gapless sector and thereby \emph{enhance} the Luttinger parameter $K_+$, leading to slower algebraic decay of single-particle correlations resulting in enhanced coherence.  We  identify the stability condition for this enhancement and discuss its physical interpretation in terms of circuit-tunable parameters.

The remainder of the paper is organized as follows.  We introduce the microscopic MCJC+Cross-Kerr Hamiltonian and the dispersive origin of the cross-Kerr interaction, then perform the polariton projection to obtain the effective bipartite EBH description.  We next develop the continuum bosonization theory, analyze the gapped $(-)$ sector and the resulting effective LL for the $(+)$ sector, and finally discuss parameter regimes and experimental relevance for superconducting-circuit implementations.

\section{Model Hamiltonian}
\label{sec:model}

We consider a one-dimensional multiconnected Jaynes--Cummings (MCJC) lattice consisting of alternating
 qubits and single-mode resonators, with each qubit coupled to its left and right neighbouring
resonators with different couplings. In addition, we engineer a nearest-neighbour density--density (cross--Kerr) interaction between
adjacent resonator modes using dispersively coupled ladder-type superconducting qutrits \cite{Liu2017}. 
Throughout we set $\hbar=1$.

\subsection{MCJC lattice}
\label{subsec:mcjc}
\begin{figure}
  \includegraphics[width=\linewidth]{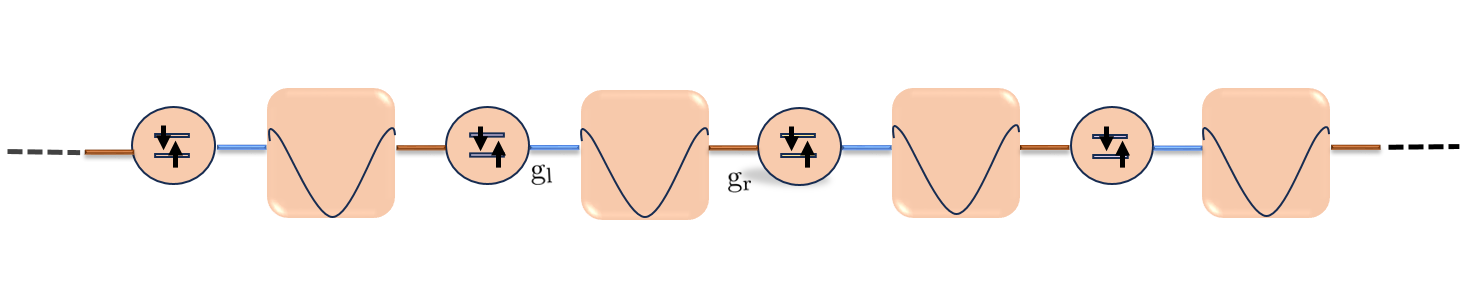}
  \caption{Schematic of a multiconnected Jaynes--Cummings (MCJC) lattice. 
  Each cavity mode is locally coupled to a qubit, while adjacent cavities are indirectly connected via qubit-mediated couplings with $g_l$ and $g_r$. }
  \label{fig:mcjc_bare}
\end{figure}

The MCJC lattice, shown in Fig.~\ref{fig:mcjc_bare}, consists of cavity modes locally coupled to qubits and indirectly connected via asymmetric qubit-mediated hopping processes. 
In this form, the system is translationally invariant and admits a polaritonic band structure at low energies\cite{Xue2017}.
We label qubits by odd integers $2i-1$ and resonators by even integers $2i$ (periodic boundary conditions
are assumed unless stated otherwise), where $i=1,2,....$ is an integer. The MCJC Hamiltonian is
\begin{align}
H_{\rm MCJC} &= H_{0}+H_{l}+H_{r},
\label{eq:Hmcjc_total}\\[4pt]
H_{0} &= \sum_{i}\left(\frac{\omega_{z}}{2}\sigma^{z}_{2i-1} + \omega_{c}\,a^{\dagger}_{2i}a_{2i}\right),
\label{eq:H0}\\[4pt]
H_{l} &= \sum_{i} g_{l}\left(\sigma^{+}_{2i-1}a_{2i-2}+a^{\dagger}_{2i-2}\sigma^{-}_{2i-1}\right),
\label{eq:Hl}\\[4pt]
H_{r} &= \sum_{i} g_{r}\left(\sigma^{+}_{2i-1}a_{2i}+a^{\dagger}_{2i}\sigma^{-}_{2i-1}\right),
\label{eq:Hr}
\end{align}
where $a_{2i}$ annihilates a photon in resonator $2i$ with frequency $\omega_c$, and $\sigma^{z,\pm}_{2i-1}$
are Pauli operators for qubit $2i-1$ with transition frequency $\omega_z$. The parameters $g_{l}$ and $g_{r}$
denote the left and right qubit--resonator couplings, respectively. We also define the qubit--resonator detuning
$\Delta \equiv \omega_z-\omega_c$.

\subsection{Resonator--resonator cross-Kerr couplings}
\label{subsec:qutrit_micro}
\begin{figure}[t]
  \includegraphics[width=\linewidth]{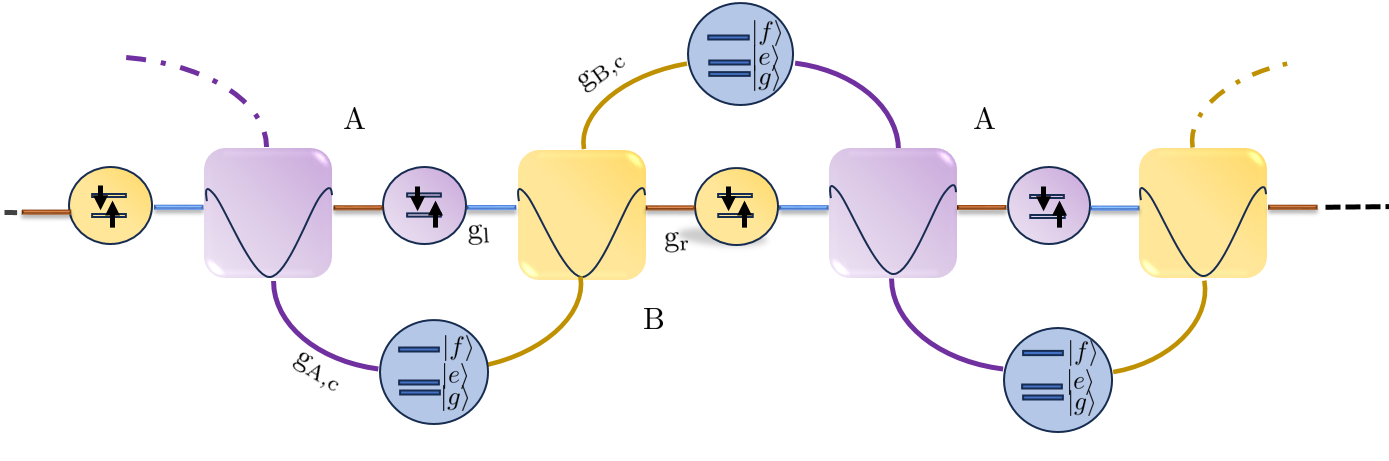}
  \caption{MCJC lattice after the installation of qutrit-mediated cross-Kerr interactions. 
  Auxiliary three-level systems couple neighboring cavities and generate effective nearest-neighbor cross-Kerr terms -$\chi\, n_i n_{i+1}$. 
  This interaction alternates between adjacent links, rendering the lattice into a bipartite system with two inequivalent sublattices $A$ and $B$.}
  \label{fig:mcjc_ck}
\end{figure}
To engineer effective interactions beyond the standard on-site one, we introduce auxiliary three-level systems that couple neighboring cavities, as illustrated in Fig.~\ref{fig:mcjc_ck}. 
These qutrits generate nearest-neighbor cross-Kerr interactions of the form $\chi\, n_i n_{i+1}$ \cite{Liu2017,PhysRevLett.115.180501,PhysRevX.10.011045}. 
Because the interaction pattern alternates between adjacent links, the resulting lattice becomes bipartite, with two inequivalent sublattices $A$ and $B$. 
As we show below, this structure naturally leads to symmetric and antisymmetric collective modes, with only the symmetric mode remaining gapless and controlling the low-energy physics.
For notational clarity, we relabel the resonator modes into
a two-site unit-cell basis $(A,B)$:
\begin{align}
a_{j,A} \equiv a_{2j-1}, \qquad a_{j,B} \equiv a_{2j},
\end{align}
so that each bond $j$ couples resonators $(j,A)$ and $(j,B)$.
On each bond $j$ we introduce a qutrit with states $\ket{g}_j,\ket{e}_j,\ket{f}_j$ and bare transition
frequencies $\omega_{eg}$ and $\omega_{fe}$ (so that $E_f = \omega_{eg}+\omega_{fe}$ in the lab frame).

The Hamiltonian is
\begin{align}
H_{\rm bond} &= \sum_{j}\left(H^{(j)}_{\rm res}+H^{(j)}_{\rm qut}+H^{(j)}_{\rm int}\right),
\label{eq:Hbond_total}
\end{align}
with
\begin{align}
H^{(j)}_{\rm res} &= \omega_{A}\,a^\dagger_{j,A}a_{j,A} + \omega_{B}\,a^\dagger_{j,B}a_{j,B},
\label{eq:Hbond_res}\\[4pt]
H^{(j)}_{\rm qut} &= \omega_{eg}\,\ket{e}_j\!\bra{e} + (\omega_{eg}+\omega_{fe})\,\ket{f}_j\!\bra{f},
\label{eq:Hbond_qut}\\[4pt]
H^{(j)}_{\rm int} &= g_{A}\left(a_{j,A}\,\sigma^{(j)}_{eg}+a^\dagger_{j,A}\,\sigma^{(j)}_{ge}\right)
+ g_{B}\left(a_{j,B}\,\sigma^{(j)}_{fe}+a^\dagger_{j,B}\,\sigma^{(j)}_{ef}\right).
\label{eq:Hbond_int}
\end{align}
Here $\sigma^{(j)}_{eg}\equiv \ket{e}_j\!\bra{g}$ and $\sigma^{(j)}_{fe}\equiv \ket{f}_j\!\bra{e}$
(and Hermitian conjugates accordingly). The couplings $g_A$ and $g_B$ are engineered such that mode $A$ couples
dispersively to the $\ket{g}\leftrightarrow\ket{e}$ transition and mode $B$ couples dispersively to the
$\ket{e}\leftrightarrow\ket{f}$ transition.

The full microscopic Hamiltonian of the lattice is then
\begin{align}
H_{\rm micro} = H_{\rm MCJC} + H_{\rm bond}.
\label{eq:Hfull_micro}
\end{align}

\subsection{Dispersive regime and elimination of the qutrit}
\label{subsec:dispersive}

We work in the dispersive regime on each bond, where the detunings
\begin{align}
\delta_A \equiv \omega_{eg}-\omega_{A},\qquad \delta_B \equiv \omega_{fe}-\omega_{B},
\end{align}
satisfy the conditions
\begin{align}
|\delta_A| \gg g_A\sqrt{\bar{n}_A+1},\qquad
|\delta_B| \gg g_B\sqrt{\bar{n}_B+1},
\label{eq:dispersive_conditions}
\end{align}
for the relevant average photon-number $\bar n_{A}$ and $\bar n_{B}$ respectively.
Under these conditions the qutrit remains near its ground state $\ket{g}$ and can be eliminated perturbatively
by applying successive Schrieffer--Wolff transformations. This results in effective resonator--only interaction on each bond.

To second order in the effective two-photon coupling, one obtains an induced cross--Kerr interaction of the form
\begin{align}
H_{XK} = -\chi\sum_{j}\,\hat n_{j,A}\,\hat n_{j,B},
\label{eq:HXK_def}
\end{align}
where $\hat n_{j,\alpha}=a^\dagger_{j,\alpha}a_{j,\alpha}$.
A convenient parametrization expresses the coupling as
\begin{align}
\chi = \frac{\lambda^2}{\Delta_c},\qquad
\lambda = \frac{g_A g_B}{2}\left(\frac{1}{|\delta_A|}+\frac{1}{\delta_B}\right),\qquad
\Delta_c = \delta_B-|\delta_A|.
\label{eq:chi_lambda_Delta}
\end{align}
In addition, dispersive elimination generates Stark shifts of the form
\begin{align}
H_{\rm Stark} = -\sum_{j}\left(\delta\omega_A\,\hat n_{j,A} + \delta\omega_B\,\hat n_{j,B}\right),
\label{eq:Hstark}
\end{align}
with 
\begin{align}
\delta\omega_A \sim \frac{g_A^2}{\delta_A},\qquad \delta\omega_B \sim \frac{g_B^2}{\delta_B},
\label{eq:stark_shifts}
\end{align}
up to small corrections.
In what follows we absorb these shifts into renormalized resonator frequencies (or, equivalently, into the chemical
potentials of the effective bosonic model), and focus on the induced density--density interaction \eqref{eq:HXK_def}.

The resulting effective lattice Hamiltonian after eliminating the qutrit degrees of freedom is therefore
\begin{align}
H_{\rm full}^{(\rm eff)} = H_{\rm MCJC} + H_{XK} + H_{\rm Stark},
\label{eq:Hfull_eff}
\end{align}
where $H_{\rm MCJC}$ is given in Eqs.~\eqref{eq:Hmcjc_total}--\eqref{eq:Hr} and $H_{XK}$ is given by
Eq.~\eqref{eq:HXK_def}.

\subsection{Mapping to a bipartite extended Bose--Hubbard model}
\label{subsec:polaritons_to_EBH}

We now rewrite the low-energy lattice Hamiltonian in the polariton representation following the standard MCJC
construction of \cite{Xue2017}. Each unit cell consists of a qubit at site $(2i-1)$ coupled to a resonator at site $(2i)$ via $g_r$,
while the intercell coupling $g_l$ connects the qubit $(2i-1)$ to its neighboring resonators. We focus on the lower
polariton branch at zero qubit--resonator detuning ($\Delta_A=\Delta_B=0$).

\paragraph*{Polariton operators and number operators.}
Let $\ket{n,-}_{j,\alpha}$ denote the \emph{lower} Jaynes--Cummings polariton eigenstate in the $n$-excitation manifold
of the local qubit--resonator subsystem labelled by site $j$ and sublattice $\alpha\in\{A,B\}$, with the identification
$A\equiv(2i-1)$ and $B\equiv(2i)$. We introduce polariton ladder operators
\begin{align}
p_{j,n,\alpha}\equiv \ket{0}_{j,\alpha}\!\bra{n,-}_{j,\alpha},\qquad
p^\dagger_{j,n,\alpha}\equiv \ket{n,-}_{j,\alpha}\!\bra{0}_{j,\alpha},
\end{align}
and define the polariton number operators
\begin{align}
\hat n_{j,\alpha} \equiv \sum_{n\ge 0} n\, p^\dagger_{j,n,\alpha}p_{j,n,\alpha}.
\end{align}

\paragraph*{Two-mode lower-polariton Hamiltonian}
Collecting all the terms, one can write the effective lower-polariton lattice Hamiltonian 
\begin{align}
H_{\rm pol} \;=\; H_A + H_B + H_{\rm hop\text{-}1}+H_{\rm hop\text{-}2}+H_{XK}.
\label{eq:Hpol_compact}
\end{align}

\noindent\textit{(i) On-site terms.} The local lower-polariton spectrum yields
\begin{align}
H_A &= \sum_i\sum_{n\ge 0}\varepsilon^A_n\,p^\dagger_{(2i-1),n,A}\,p_{(2i-1),n,A}\;-\;\delta\omega_A\sum_i \hat n_{(2i-1),A},
\label{eq:HA_compact}\\[2pt]
H_B &= \sum_i\sum_{m\ge 0}\varepsilon^B_m\,p^\dagger_{(2i),m,B}\,p_{(2i),m,B},
\label{eq:HB_compact}
\end{align}
where $\varepsilon^{A,B}_n$ are the lower-polariton eigenenergies for the local Jaynes--Cummings subsystems, and
$\delta\omega_A$ denotes the (small) dispersive Stark shift generated by the qutrit coupler on the $A$-site resonator
mode (which we shall ignore).

\noindent\textit{(ii) Intercell transfer terms.} The qubit-mediated coupling $g_l$ induces transfer or hopping between neighboring
lower-polariton states. It is convenient to express these hopping processes through coefficient matrices multiplying
universal polariton-transfer operators:
\begin{align}
H_{\rm hop\text{-}1} &= -g_l\sum_i\sum_{n,m\ge 0} \,T^{(1)}_{nm}\Big(\mathcal V^{(1)}_{i;nm}+\mathrm{H.c.}\Big),
\label{eq:Hhop1_compact}\\[2pt]
H_{\rm hop\text{-}2} &= -g_l\sum_i\sum_{n,m\ge 0} \,T^{(2)}_{nm}\Big(\mathcal V^{(2)}_{i;nm}+\mathrm{H.c.}\Big),
\label{eq:Hhop2_compact}
\end{align}
where $\mathcal V^{(1)}_{i;nm}$ transfers an excitation between the intra-cell pair $((2i-1),A)\leftrightarrow (2i),B$,
and $\mathcal V^{(2)}_{i;nm}$ transfers between the inter-cell pair $(2i),B \leftrightarrow (2i+1),A$.
Explicitly, these operators have the generic polariton form
\begin{align}
\mathcal V^{(1)}_{i;nm} &\equiv
p^\dagger_{(2i-1),n-1,A}\,p_{(2i-1),n,A}\;
p^\dagger_{(2i),m+1,B}\,p_{(2i),m,B},
\label{eq:V1_def}\\[2pt]
\mathcal V^{(2)}_{i;nm} &\equiv
p^\dagger_{(2i),m-1,B}\,p_{(2i),m,B}\;
p^\dagger_{(2i+1),n+1,A}\,p_{(2i+1),n,A}.
\label{eq:V2_def}
\end{align}
The coefficients $T^{(1)}_{nm}$ and $T^{(2)}_{nm}$ encode the appropriate square-root factors and lower-polariton mixing
coefficients that result from the Jaynes--Cummings eigenstates; their explicit expressions are collected in
Appendix~\ref{app:polariton_hopping}.

\noindent\textit{(iii) Cross--Kerr term.} The dispersively engineered qutrit couplers generate an attractive density--density
interaction between neighboring resonator modes, which in the lower-polariton sector becomes
\begin{align}
H_{XK} = -\chi\sum_i \hat n_{(2i-1),A}\,\hat n_{(2i),B}\;-\;\chi\sum_i \hat n_{(2i),B}\,\hat n_{(2i+1),A}.
\label{eq:HXK_compact}
\end{align}

\paragraph*{Reduction to a bipartite extended Bose--Hubbard form.}
We now restrict to the low-energy subspace of the lower polaritons and truncate the local occupation to the two-excitation
manifold ($n\le 2$), so that the local anharmonic spectrum defines an effective on-site repulsion $U$, while the transfer
terms \eqref{eq:Hhop1_compact}--\eqref{eq:Hhop2_compact} define effective intra- and inter-cell hopping amplitudes.
Moreover, the residual on-site energy differences (from the Stark shift), although can be absorbed into staggered chemical
potentials $\mu_A$ and $\mu_B$ have been assumed to be small enough to be ignored. This yields the bipartite extended Bose--Hubbard-like Hamiltonian
\begin{align}
H &= H_t+H_U+H_\chi+H_\mu,
\label{eq:HEBH_again}\\[2pt]
H_t &= -\sum_j\Big(t_{\rm intra}\,b^\dagger_{j,A}b_{j,B}+t_{\rm inter}\,b^\dagger_{j,B}b_{j+1,A}+\mathrm{H.c.}\Big),\\[2pt]
H_U &= \frac{U}{2}\sum_{j,\alpha\in\{A,B\}} n_{j,\alpha}\big(n_{j,\alpha}-1\big),\\[2pt]
H_\chi &= -\chi\sum_{\langle (j,\alpha),(j',\beta)\rangle} n_{j,\alpha}\,n_{j',\beta},\\[2pt]
H_\mu &= -\sum_j\left(\mu_A\,n_{j,A}+\mu_B\,n_{j,B}\right),
\end{align}
which is the starting point for the continuum bosonization analysis below \cite{PhysRevB.61.12474,Ejima2011PRL}. It is to be noted that for our system $t_{inter}=t_{intra}=g_l/4$.

\section{Low-energy continuum description}
\label{sec:continuum}

Starting from the bipartite extended Bose--Hubbard Hamiltonian
[Eqs.~\eqref{eq:HEBH_again}], we now derive a low-energy continuum LL model
using the harmonic-fluid (bosonization) approach\,\cite{Haldane1981JPC,GiamarchiBook2003}. Our goal is to identify the
collective modes that govern long-range behaviour of the system, determine the role of the
attractive cross--Kerr interaction, and obtain the effective LL
description for correlation functions.

For each sublattice $\alpha\in\{A,B\}$ we introduce continuum bosonic fields
$\phi_\alpha(x)$ and $\theta_\alpha(x)$ through the standard long-wavelength
representation
\begin{align}
b_\alpha(x) &\simeq \sqrt{\rho_{0,\alpha}+\delta\rho_\alpha(x)}\,e^{i\theta_\alpha(x)},\\
\delta\rho_\alpha(x) &\simeq -\frac{1}{\pi}\partial_x\phi_\alpha(x),
\end{align}
supplemented, by the leading density harmonic
\begin{align}
\rho_\alpha(x)=\rho_0-\frac{1}{\pi}\partial_x\phi_\alpha(x)
+ A_1\cos\!\left(2\pi\rho_0 x-2\phi_\alpha(x)\right)+\cdots .
\end{align}
The canonical commutation relations are
\begin{align}
[\phi_\alpha(x),\partial_y\theta_\beta(y)]
=i\pi\,\delta_{\alpha\beta}\delta(x-y).
\end{align}

It is convenient to introduce symmetric and antisymmetric combinations on the
two-site unit cell,
\begin{align}
\phi_\pm=\frac{\phi_A\pm\phi_B}{\sqrt{2}},\qquad
\theta_\pm=\frac{\theta_A\pm\theta_B}{\sqrt{2}},
\end{align}
which diagonalize the Gaussian sector at long wavelengths. The $(+)$ fields
describe in-phase density and phase fluctuations of a unit cell, while the
$(-)$ fields describe relative (out-of-phase) fluctuations between the two
sublattices.

We now summarize the leading non-oscillatory continuum contributions generated
by the lattice Hamiltonian; details of the derivation are provided in
Appendix~\ref{app:continuum}. The hopping processes generate both
phase-gradient terms and a Josephson-like mass for the relative phase,
\begin{align}
H_t^{\rm cont}\supset
\int dx\Bigg[
\frac{M}{2}\,\theta_-^2
+\frac{s_+}{2}(\partial_x\theta_+)^2
+\frac{s_-}{2}(\partial_x\theta_-)^2
+\lambda(\partial_x\theta_+)(\partial_x\theta_-)
\Bigg],
\label{tlattice}
\end{align}
where $M$ is generated by the strong intra-cell coupling, $s_\pm$ are phase
stiffnesses proportional to the hopping amplitude and average density, and
$\lambda$ denotes a residual gradient coupling. The mass term pins $\theta_-$
and gaps the antisymmetric sector, while $\theta_+$ remains gapless. Their dependence on the Bose-Hubbard parameters are shown in the Appendix~\ref{app:continuum}, section 1.

The on-site interaction yields the standard compressibility contribution,
\begin{align}
H_U^{\rm cont}
=\frac{Ua}{2\pi^2}\int dx\left[
(\partial_x\phi_+)^2+(\partial_x\phi_-)^2
\right],
\label{ulattice}
\end{align}
while a staggered chemical potential $\mu_A\neq\mu_B$ produces a total-derivative
(tilt) term for $\phi_-$ and, at special commensurate fillings, a non-oscillatory
pinning cosine.

The attractive nearest-neighbor cross--Kerr interaction modifies the Gaussian
sector by shifting the compressibilities of the symmetric and antisymmetric
modes with opposite signs,
\begin{align}
H_\chi^{\rm cont}
= -\frac{\chi a}{2\pi^2}\int dx\,(\partial_x\phi_+)^2
+ \frac{\chi a}{2\pi^2}\int dx\,(\partial_x\phi_-)^2,
\label{xlattice}
\end{align}
and additionally generates nonlinear cosine operators through products of
density harmonics.

Collecting the leading \emph{non-oscillatory} quadratic contributions from the
hopping and interaction sectors, we obtain the full two-mode Gaussian continuum
Hamiltonian
\begin{align}
H_{\rm cont} &=
\int dx \Bigg[
\frac{M}{2}\,\theta_-^2
+\frac{s_+}{2}\big(\partial_x\theta_+\big)^2
+\frac{s_-}{2}\big(\partial_x\theta_-\big)^2
+\lambda\,\big(\partial_x\theta_+\big)\big(\partial_x\theta_-\big)
\Bigg]
\nonumber\\
&\quad
+\int dx \Bigg[
\frac{\kappa_+}{2}\big(\partial_x\phi_+\big)^2
+\frac{\kappa_-}{2}\big(\partial_x\phi_-\big)^2
\Bigg],
\label{eq:Hcont_full}
\end{align}
with compressibility coefficients
\begin{align}
\kappa_\pm=\frac{a}{2\pi^2}\big(U\mp 2\chi\big).
\end{align}
Thus the attractive cross--Kerr interaction reduces the symmetric-mode
compressibility while enhancing that of the antisymmetric mode.

Nonlinear cosine operators arise from the periodic density harmonics and from
products of such harmonics when interactions couple nearby sites. In particular,
the leading non-oscillatory pinning term generated by the cross--Kerr interaction
in the antisymmetric sector takes the form
\begin{align}
- g_{\chi,-}\int dx\,\cos\!\big(2\sqrt{2}\,\phi_-(x)\big),
\end{align}
with a nonuniversal amplitude $g_{\chi,-}\propto \chi A_1^2$. Cosine operators
involving $\phi_+$ are oscillatory and average to zero at generic incommensurate
fillings, becoming relevant only at special commensurate densities. The antisymmetric sector can therefore be gapped either by the Josephson mass,
which pins $\theta_-$, or by a relevant $\phi_-$ cosine, which pins $\phi_-$. In
both cases a finite gap $\Delta_-$ is generated, and for energies well below this
gap the antisymmetric sector may be integrated out. Carrying out this integration yields an effective single-component Luttinger liquid for the symmetric mode,
\begin{align}
H_{\rm eff,+}
=\frac{u_+}{2\pi}\int dx\left[
K_+(\pi\Pi_+)^2+\frac{1}{K_+}(\partial_x\phi_+)^2
\right],
\end{align}
with $\Pi_+=\partial_x\theta_+/\pi$. To leading order,
\begin{align}
K_+\simeq \sqrt{\frac{s_+}{\kappa_+}},\qquad
\kappa_+\propto (U-2\chi),
\end{align}
so that the attractive cross--Kerr interaction increases $K_+$. Stability of the Gaussian theory requires
$U-2\chi>0$. As a result, the single-particle correlation function decays as
\begin{align}
G_1(x)\sim |x|^{-1/(4K_+)},
\end{align}
demonstrating that increasing $\chi$ leads to slower decay of coherence and
stronger quasi-long-range order. In the following we focus on incommensurate
fillings, for which symmetric-sector umklapp processes are absent\,\cite{Haldane1981JPC,GiamarchiBook2003}.

\section{Results and interpretations}
\label{sec:results}

We now analyze the spatial correlation functions that characterize the low-energy steady state of the system in the Luttinger-liquid (LL) regime.  Our focus is on three complementary observables: the first-order coherence $g^{(1)}(x)$, the spatial density-density correlation $g^{(2)}(x)$, and the connected qubit magnetization correlator $C_{zz}(x)$.  These probe, respectively, phase coherence, density fluctuations, and the qubit sector, providing a comprehensive test of the continuum bosonized description developed in Sec.~\ref{sec:continuum}.  Throughout this section we restrict attention to distances $x\gtrsim a$, where the LL description is valid.

\subsection*{First-order coherence and enhancement of quasi-long-range order}

\begin{figure}[t]
  \centering
  \includegraphics[width=\linewidth]{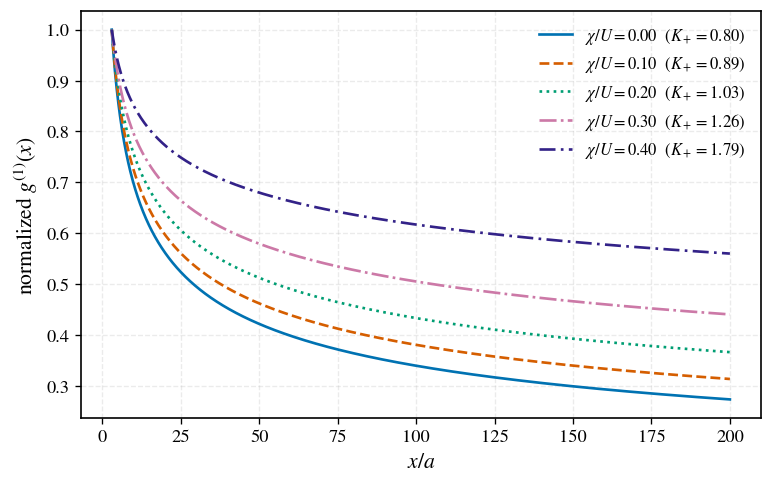}
  \caption{First-order coherence $g^{(1)}(x)$ as a function of distance $x/a$ for different values of the cross-Kerr interaction $\chi/U$. The algebraic decay becomes progressively slower with increasing $\chi/U$, reflecting an increase of the Luttinger parameter $K_+$ and enhanced quasi-long-range phase coherence.}
  \label{fig:g1}
\end{figure}

We begin with the normalized first-order coherence function
\begin{equation}
g^{(1)}(x)
\equiv
\frac{\langle a^\dagger(x)\, a(0)\rangle}{\langle a^\dagger a\rangle},
\end{equation}
shown in Fig.~\ref{fig:g1}.  Within Luttinger-liquid theory, single-particle coherence is governed by fluctuations of the phase field $\theta_+(x)$ associated with the gapless symmetric mode.  Bosonization yields the asymptotic form
\begin{equation}
g^{(1)}(x)
\sim
\left(\frac{a}{|x|}\right)^{\frac{1}{4K_+}},
\qquad (x \gg a),
\label{eq:g1_LL}
\end{equation}
where $K_+$ is the Luttinger parameter and $a$ is a short-distance cutoff \cite{Haldane1981JPC,GiamarchiBook2003}.  

As seen in Fig.~\ref{fig:g1}, $g^{(1)}(x)$ exhibits a clear algebraic decay for all values of the cross-Kerr interaction $\chi/U$, consistent with quasi-long-range order in one dimension.  Importantly, the decay becomes progressively slower as $\chi/U$ increases.  This reflects an increase of the Luttinger parameter $K_+$, indicating that the attractive cross-Kerr interaction effectively softens the interactions in the symmetric channel \cite{PhysRevLett.111.053601}.  Physically, this corresponds to an enhancement of phase coherence, while remaining consistent with the absence of true long-range order in one dimension.

\subsection*{Spatial density-density correlations and separation of phase and density sectors}

\begin{figure}[t]
  \centering
  \includegraphics[width=\linewidth]{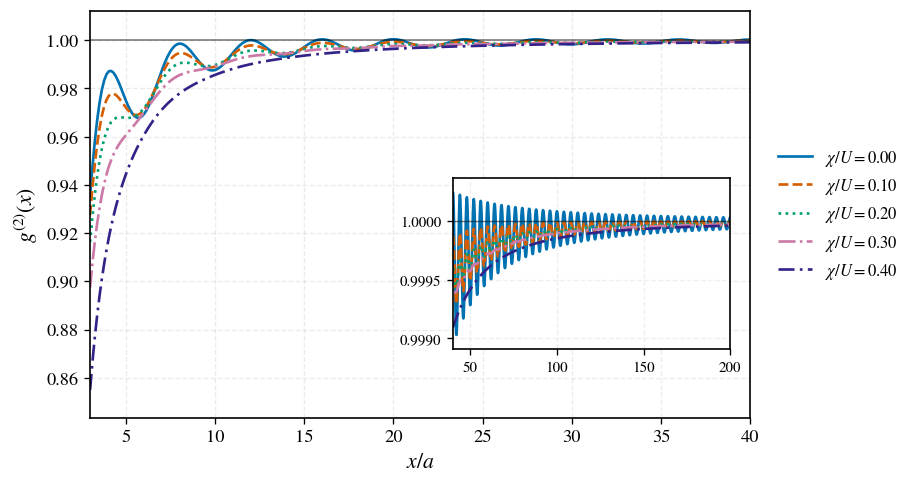}
  \caption{Spatial density-density correlations $g^{(2)}(x)$ versus $x/a$ for different $\chi/U$. All curves rapidly approach the uncorrelated value $g^{(2)}(x)=1$, with weak oscillatory corrections. The inset highlights the asymptotic convergence at large distances, consistent with the Luttinger-liquid prediction.}
  \label{fig:g2}
\end{figure}

We next consider the spatial density-density correlation function,
\begin{equation}
g^{(2)}(x)
\equiv
\frac{\langle n(x)\, n(0)\rangle}{\langle n\rangle^2},
\qquad
n(x)=a^\dagger(x)a(x),
\qquad
x\neq0
\end{equation}
shown in Fig.~\ref{fig:g2}.  Writing $n(x)=\rho_0+\delta n(x)$, where $\rho_0$ is the mean density, this can be expressed as
\begin{equation}
g^{(2)}(x)
=
1+
\frac{\langle \delta n(x)\delta n(0)\rangle}{\rho_0^2}.
\end{equation}

In Luttinger-liquid theory, the density--density correlator has the universal long-distance form
\begin{equation}
\langle \delta n(x)\delta n(0)\rangle
\simeq
-\frac{K_+}{2\pi^2}\frac{1}{x^2}
+
A_{2k}
\frac{\cos(2\pi\rho_0 x)}{|x|^{2K_+}}
+\cdots ,
\label{eq:density_corr}
\end{equation}
where $A_{2k}$ is a nonuniversal amplitude.  Substituting into the definition of $g^{(2)}(x)$ yields
\begin{equation}
g^{(2)}(x)
\simeq
1
-\frac{K_+}{2\pi^2\rho_0^2}\frac{1}{x^2}
+
\frac{A_{2k}}{\rho_0^2}
\frac{\cos(2\pi\rho_0 x)}{|x|^{2K_+}}.
\label{eq:g2_LL}
\end{equation}

Figure~\ref{fig:g2} shows that $g^{(2)}(x)$ rapidly approaches the uncorrelated value $g^{(2)}=1$ with increasing distance, with only weak oscillatory corrections at intermediate scales.  The inset highlights the asymptotic convergence at large $x$.  The qualitative contrast between Figs.~\ref{fig:g1} and \ref{fig:g2} is a hallmark of Luttinger-liquid physics: phase coherence decays slowly due to gapless phase fluctuations, density correlations are comparatively short-ranged and controlled by the universal $1/x^2$ term in Eq.~\eqref{eq:density_corr}.  This clear separation of phase and density sectors is a property of the Luttinger-liquid regime.

\subsection*{Qubit magnetization correlations and inheritance of Luttinger-liquid scaling}

\begin{figure}[H]
  \includegraphics[width=\linewidth]{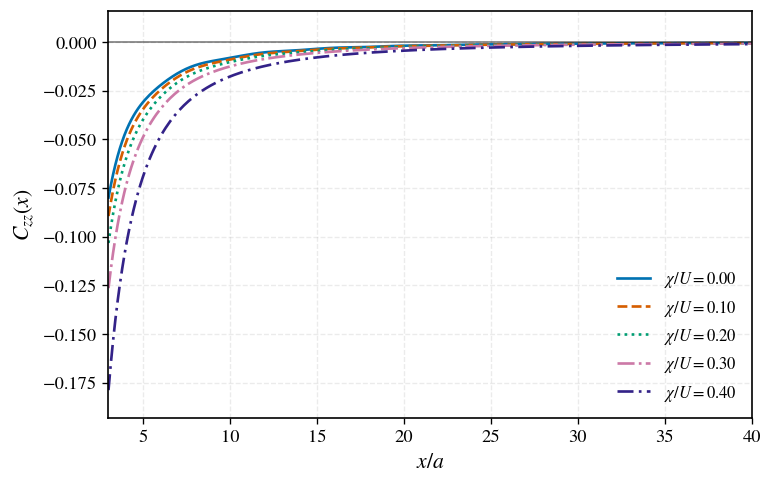}
  \caption{Connected qubit magnetization correlator $C_{zz}(x)$ as a function of distance $x/a$. The correlator decays algebraically to zero for all $\chi/U$. Variations with $\chi/U$ reflect nonuniversal amplitudes, while the asymptotic scaling is governed by the same Luttinger parameter $K_+$ as the photonic correlators.}
  \label{fig:Czz}
\end{figure}

Finally, we analyze the connected qubit magnetization correlator
\begin{equation}
C_{zz}(x)
\equiv
\langle \sigma^z(x)\sigma^z(0)\rangle
-
\langle \sigma^z\rangle^2,
\end{equation}
shown in Fig.~\ref{fig:Czz}.  In the low-energy manifold relevant to the LL regime, the local qubit operator $\sigma^z(x)$ can be expanded in terms of the bosonized fields as
\begin{equation}
\sigma^z(x)
=
c_0
+
c_1\,\partial_x\phi_+(x)
+
c_2
\cos\!\bigl(2\pi\rho_0 x - 2\phi_+(x)\bigr)
+\cdots ,
\label{eq:sigmaz_expansion}
\end{equation}
where $\phi_+(x)$ is the density field of the symmetric mode and the coefficients $c_i$ are nonuniversal.

This expansion implies the asymptotic form
\begin{equation}
C_{zz}(x)
\simeq
-\frac{\mathcal A(\chi)}{x^2}
+
\mathcal B(\chi)
\frac{\cos(2\pi\rho_0 x)}{|x|^{2K_+}}
+\cdots ,
\label{eq:Czz_LL}
\end{equation}
where $\mathcal A(\chi)$ and $\mathcal B(\chi)$ are nonuniversal amplitudes, while the exponents are fixed by the same Luttinger parameter $K_+$ that governs the photonic correlations \cite{Haldane1981JPC,GiamarchiBook2003}.

As shown in Fig.~\ref{fig:Czz}, $C_{zz}(x)$ decays algebraically to zero for all values of $\chi/U$, consistent with the absence of long-range magnetic order.  The systematic dependence on $\chi/U$ at short distances reflects changes in nonuniversal amplitudes, while the asymptotic behavior is fully consistent with the Luttinger-liquid prediction in Eq.~\eqref{eq:Czz_LL}.  This demonstrates that the qubit subsystem inherits the same low-energy Luttinger-liquid physics as the photonic degrees of freedom, providing a nontrivial cross-channel consistency check of the continuum description.

Taken together, the correlation functions presented in Figs.~\ref{fig:g1}, \ref{fig:g2}, and \ref{fig:Czz} establish a coherent physical picture of a gapless one-dimensional Luttinger liquid in the symmetric channel, with an interaction-dependent Luttinger parameter $K_+$ tunable via the cross-Kerr coupling $\chi$. The enhancement of quasi-long-range phase coherence, the rapid saturation of spatial density-density correlations, and the algebraic decay of qubit magnetization correlations are all quantitatively consistent with this description. Additional scaling diagnostics and channel-equivalence checks are presented in the Appendix.

\section{Conclusion}

We have shown that engineered nearest-neighbor cross-Kerr interactions in a multiconnected Jaynes--Cummings lattice provide a route to enhancing long-distance coherence in one-dimensional circuit-QED systems. Starting from a microscopic circuit realization with dispersively coupled qutrits, the resulting effective model admits a projection onto the lower-polaritonic manifold, where it maps onto an extended Bose--Hubbard chain with on-site repulsion and tunable attractive density--density interactions. This establishes a direct correspondence between experimentally accessible circuit parameters and the low-energy collective modes of interacting bosons.

A central outcome of our analysis is that the bipartite structure of the MCJC lattice  separates the density fluctuations and phases of the Luttinger liquid into symmetric and antisymmetric sectors. The antisymmetric mode acquires a finite gap, allowing it to be integrated out and leaving a single gapless Luttinger liquid governing the long-distance behavior. Within this regime, the attractive cross-Kerr interaction renormalizes the effective compressibility of the remaining mode without destabilizing the system, provided the stability condition $U-2\chi>0$ is satisfied. As a result, the Luttinger parameter $K_+$ increases monotonically with the strength of the cross-Kerr coupling, leading to a parametrically slower algebraic decay of single-particle coherence. In this sense, the cross-Kerr interaction acts as a genuine coherence-enhancement knob rather than merely an additional perturbation.

The universal long-distance correlations of both photonic and qubit observables are governed by the same low-energy theory here, demonstrating that the qubit sector faithfully inherits the collective Luttinger-liquid physics of the polaritonic degrees of freedom. This unification highlights the MCJC platform as a particularly clean realization of an interacting quantum fluid, in which microscopic hybridization and strong light--matter coupling do not obscure the emergent continuum description.

Our results place this cross-Kerr--engineered circuit-QED lattice within the broader universality class of attractive one-dimensional bosonic systems, while offering a level of tunability and control that is difficult to achieve in other platforms. Beyond the parameter regime considered here, commensurate fillings, stronger interactions, or competing gap-opening mechanisms are expected to generate departures from the single-mode Luttinger-liquid picture, potentially stabilizing insulating or density-ordered phases. More broadly, gap-opening mechanisms tied to constrained dynamics—such as dipole physics in tilted Bose–Hubbard chains and recent work on fractonic Luttinger liquids—suggest additional directions for engineered one-dimensional platforms \cite{PhysRevB.107.195131,PhysRevLett.120.195301,PhysRevLett.132.071603}. Exploring these regimes, as well as benchmarking the present predictions against exact numerics and experimentally accessible correlation measurements, constitutes a natural direction for future work.

\section{Acknowledgement}
NS is grateful to University Grants Commission (UGC), Govt. of India, for support.

\appendix

\section{Polariton projection and hopping-coefficient matrices}
\label{app:polariton_hopping}

This appendix collects the polariton projection formulas and the coefficient matrices
entering the polaritonic hopping terms in the main-text lattice Hamiltonian.

\subsection{Local JC eigenstates and projected operators}
For each sublattice $\alpha\in\{A,B\}$, the local Jaynes--Cummings eigenstates in the $n$-excitation manifold are
\begin{equation}
\ket{n,+}_\alpha = \gamma_{n,\alpha}\ket{n,g}_\alpha + \rho_{n,\alpha}\ket{n-1,e}_\alpha,\qquad
\ket{n,-}_\alpha = \rho_{n,\alpha}\ket{n,g}_\alpha - \gamma_{n,\alpha}\ket{n-1,e}_\alpha,
\end{equation}
with $\gamma_{n,\alpha}^2+\rho_{n,\alpha}^2=1$. Projecting onto the lower polariton branch,
the photon and qubit lowering operators admit the expansions
\begin{align}
a_{i,\alpha}
&\simeq \sum_{n\ge 1} t_{n,\alpha}\;
p^\dagger_{i,n-1,\alpha}p_{i,n,\alpha},\\
\sigma^-_{i,\alpha}
&\simeq \sum_{n\ge 1} k_{n,\alpha}\;
p^\dagger_{i,n-1,\alpha}p_{i,n,\alpha},
\end{align}
where the JC matrix elements are
\begin{align}
t_{n,\alpha}
&= \sqrt{n}\,\gamma_{n,\alpha}\gamma_{n-1,\alpha}
     + \sqrt{n-1}\,\rho_{n,\alpha}\rho_{n-1,\alpha},\\
k_{n,\alpha}
&= \rho_{n,\alpha}\gamma_{n-1,\alpha}.
\end{align}

\subsection{Universal transfer operators and coefficient matrices}
In the bipartite polariton basis, the intercell transfer operators can be written in universal form
\begin{align}
V^{(1)}_{i;nm} &\equiv
p^\dagger_{(2i-1),n-1,A}\,p_{(2i-1),n,A}\;
p^\dagger_{(2i),m+1,B}\,p_{(2i),m,B},\\
V^{(2)}_{i;nm} &\equiv
p^\dagger_{(2i),m-1,B}\,p_{(2i),m,B}\;
p^\dagger_{(2i+1),n+1,A}\,p_{(2i+1),n,A}.
\end{align}
The corresponding coefficient matrices $T^{(1)}_{nm}$ and $T^{(2)}_{nm}$ depend on the microscopic
coupling channel. For qubit-assisted transfer,
\begin{equation}
T^{(1)}_{nm} = k_{n,A}\,k_{m+1,B},\qquad
T^{(2)}_{nm} = k_{m,B}\,k_{n+1,A},
\end{equation}
while for photon-assisted transfer the corresponding leg(s) are replaced as
$k_{n,\alpha}\to t_{n,\alpha}$, e.g.\ for a mixed photon--qubit process,
\begin{equation}
T^{(1)}_{nm} = k_{n,A}\,t_{m+1,B},\qquad
T^{(2)}_{nm} = t_{m,B}\,k_{n+1,A}.
\end{equation}

\section{Cross-Kerr implementation}
\label{app:crosskerr_impl}

We set $\hbar=1$. We take bosonic modes $a,b$ that satisfy $[a,a^\dagger]=[b,b^\dagger]=1$ and commute with the qutrit operators. The qutrit basis is $\{|g\rangle,|e\rangle,|f\rangle\}$.

\vspace{4pt}
In the interaction picture, the coupling Hamiltonian is
\begin{equation}
H_I(t)
= g_{A,c}\big(e^{i\delta_a t}a\,\sigma_{eg}^+ + e^{-i\delta_a t}a^\dagger\sigma_{eg}^-\big)
+ g_{B,c}\big(e^{i\delta_b t}b\,\sigma_{fe}^+ + e^{-i\delta_b t}b^\dagger\sigma_{fe}^-\big),
\end{equation}
where $\sigma_{eg}^+=|e\rangle\langle g|,\ \sigma_{eg}^-=|g\rangle\langle e|$, $\sigma_{fe}^+=|f\rangle\langle e|,\ \sigma_{fe}^-=|e\rangle\langle f|$.

\bigskip

Working in the Schr\"odinger picture and splitting $H=H_0+V$, we perform a first Schrieffer--Wolff transformation to eliminate the $\ket{g}\leftrightarrow\ket{e}$ and $\ket{e}\leftrightarrow\ket{f}$ resonant terms:
\begin{align}
H_0&=\omega_a a^\dagger a+\omega_b b^\dagger b+\omega_g|g\rangle\langle g|+\omega_e|e\rangle\langle e|+\omega_f|f\rangle\langle f|,\\
V&= g_{A,c}\,(a\sigma_{eg}^+ + a^\dagger\sigma_{eg}^-)
   + g_{B,c}\,(b\sigma_{fe}^+ + b^\dagger\sigma_{fe}^-).
\end{align}
Choosing the anti-Hermitian generator
\begin{equation}
S \;=\; -\frac{g_{A,c}}{\delta_a}\big(a\,\sigma_{eg}^+ - a^\dagger\sigma_{eg}^-\big)
      -\frac{g_{B,c}}{\delta_b}\big(b\,\sigma_{fe}^+ - b^\dagger\sigma_{fe}^-\big),
\end{equation}
we verify $[H_0,S]=-V$ using
\begin{equation}
[H_0,\,a\sigma_{eg}^+] = \delta_a\,a\sigma_{eg}^+,\qquad
[H_0,\,a^\dagger\sigma_{eg}^-] = -\delta_a\,a^\dagger\sigma_{eg}^-,
\end{equation}
and similarly for $b,\sigma_{fe}^\pm$ with $\delta_b$. Hence
\begin{equation}
[H_0,S]
= -\frac{g_{A,c}}{\delta_a}\big(\delta_a a\sigma_{eg}^+ + \delta_a a^\dagger\sigma_{eg}^-\big)
  -\frac{g_{B,c}}{\delta_b}\big(\delta_b b\sigma_{fe}^+ + \delta_b b^\dagger\sigma_{fe}^-\big)
= -V.
\end{equation}

\medskip
To second order,
\begin{equation}
H_{\rm eff} \approx H_0 + \tfrac12[S,V] + O(V^3).
\end{equation}
Split $S=S_{A}+S_{B}$ and $V=V_{A}+V_{B}$ with
\begin{align}
S_{A}&=-\frac{g_{A,c}}{\delta_a}(X-Y),\quad X\equiv a\sigma_{eg}^+,\ Y\equiv a^\dagger\sigma_{eg}^-,\\
S_{B}&=-\frac{g_{B,c}}{\delta_b}(C-D),\quad C\equiv b\sigma_{fe}^+,\ D\equiv b^\dagger\sigma_{fe}^-,\\
V_{A}&=g_{A,c}(X+Y),\qquad V_{B}=g_{B,c}(C+D).
\end{align}

Computing the contributions:
\begin{equation}
[S_{A},V_{A}] = -\frac{g_{A,c}^2}{\delta_a}\big[(X-Y),(X+Y)\big] = -\frac{2g_{A,c}^2}{\delta_a}[X,Y],
\end{equation}
and
\begin{align}
[X,Y] &= a\sigma_{eg}^+\,a^\dagger\sigma_{eg}^- - a^\dagger\sigma_{eg}^-\,a\sigma_{eg}^+ \nonumber\\
&= a a^\dagger\,|e\rangle\langle e| - a^\dagger a\,|g\rangle\langle g| \nonumber\\
&= \big(a^\dagger a + 1\big)|e\rangle\langle e| - a^\dagger a\,|g\rangle\langle g|.
\end{align}
Thus
\begin{equation}
\tfrac12[S_{A},V_{A}] = -\frac{g_{A,c}^2}{\delta_a}\Big\{a^\dagger a\big(|e\rangle\langle e|-|g\rangle\langle g|\big) + |e\rangle\langle e|\Big\}.
\end{equation}

Identical algebra with $(b,g_{B,c},f,e)$ yields
\begin{equation}
\tfrac12[S_{B},V_{B}] = -\frac{g_{B,c}^2}{\delta_b}\Big\{b^\dagger b\big(|f\rangle\langle f|-|e\rangle\langle e|\big) + |f\rangle\langle f|\Big\},
\end{equation}
equivalently
\begin{equation}
-\frac{g_{B,c}^2}{\delta_b}\big(b^\dagger b\,|e\rangle\langle e| - b b^\dagger\,|f\rangle\langle f|\big).
\end{equation}

Computing the cross-commutators:
\begin{equation}
[S_{A},V_{B}] = -\frac{g_{A,c}g_{B,c}}{\delta_a}\big[(X-Y),(C+D)\big],
\end{equation}
and evaluating nonzero elementary commutators we get
\begin{equation}
[X,C] = [a\sigma_{eg}^+,\,b\sigma_{fe}^+] = -ab\,\sigma_{fg}^+,\qquad
[Y,D] = [a^\dagger\sigma_{eg}^-,\,b^\dagger\sigma_{fe}^-] = a^\dagger b^\dagger\,\sigma_{fg}^-,
\end{equation}
(all other combinations vanish by qutrit orthogonality). Hence
\begin{equation}
[S_{A},V_{B}] = \frac{g_{A,c}g_{B,c}}{\delta_a}\big(ab\,\sigma_{fg}^+ + a^\dagger b^\dagger\,\sigma_{fg}^-\big).
\end{equation}
Similarly,
\begin{equation}
[S_{B},V_{A}] = -\frac{g_{A,c}g_{B,c}}{\delta_b}\big(ab\,\sigma_{fg}^+ + a^\dagger b^\dagger\,\sigma_{fg}^-\big).
\end{equation}
Thus
\begin{equation}
\tfrac12\big([S_{A},V_{B}]+[S_{B},V_{A}]\big) =
\frac{g_{A,c}g_{B,c}}{2}\Big(\frac{1}{\delta_a}-\frac{1}{\delta_b}\Big)\big(ab\,\sigma_{fg}^+ + a^\dagger b^\dagger\,\sigma_{fg}^-\big).
\end{equation}

\medskip
Moving to the interaction picture (restoring time dependence) the two-photon operator becomes
\begin{equation}
\lambda\big(e^{-i\Delta t}a^\dagger b^\dagger\,\sigma_{fg}^- + e^{i\Delta t}ab\,\sigma_{fg}^+\big),
\end{equation}
with frequency $\Delta=\delta_b-|\delta_a|$. From the commutator result above one finds
\begin{equation}
\lambda \simeq \frac{g_{A,c}g_{B,c}}{2}\Big(\frac{1}{\delta_a}-\frac{1}{\delta_b}\Big).
\end{equation}
Using $\delta_a=-|\delta_a|$ (since typically $\delta_a<0$) we rewrite
\begin{equation}
\frac{1}{\delta_a}-\frac{1}{\delta_b} = -\Big(\frac{1}{|\delta_a|}+\frac{1}{\delta_b}\Big),
\end{equation}
so absorbing the sign into a phase convention we define the positive coefficient
\begin{equation}
\lambda \equiv \frac{g_{A,c}g_{B,c}}{2}\Big(\frac{1}{|\delta_a|}+\frac{1}{\delta_b}\Big).
\end{equation}

\bigskip
Collecting all second-order pieces (dropping the fast free part $H_0$ in the interaction picture),
\begin{equation}
\begin{aligned}
H_e &= -\frac{g_{A,c}^2}{\delta_a}\big(a^\dagger a\,|g\rangle\langle g| - a a^\dagger\,|e\rangle\langle e|\big)
- \frac{g_{B,c}^2}{\delta_b}\big(b^\dagger b\,|e\rangle\langle e| - b b^\dagger\,|f\rangle\langle f|\big)\\
&\qquad\qquad + \lambda\big(e^{-i\Delta t}a^\dagger b^\dagger\,\sigma_{fg}^- + e^{i\Delta t}ab\,\sigma_{fg}^+\big).
\end{aligned}
\end{equation}

\medskip
Now for the second SW elimination: we remove the off-resonant $|g\rangle\leftrightarrow|f\rangle$ coupling and obtain the $\chi$ (cross-Kerr) term.

Defining operators
\begin{equation}
A \equiv a^\dagger b^\dagger\,\sigma_{fg}^-,\qquad A^\dagger = a b\,\sigma_{fg}^+,
\end{equation}
the two-photon interaction is $V(t)=\lambda(e^{-i\Delta t}A + e^{i\Delta t}A^\dagger)$. For large detuning $\Delta$ (compared to $\lambda$ and Stark shifts) we perform a second SW elimination with generator
\begin{equation}
S' = -\frac{\lambda}{\Delta}\big(e^{-i\Delta t}A - e^{i\Delta t}A^\dagger\big).
\end{equation}
The second-order correction is $\tfrac12[S',V]$. Computing the commutator:
\begin{align}
[S',V] &= -\frac{\lambda^2}{\Delta}\Big[(e^{-i\Delta t}A-e^{i\Delta t}A^\dagger),(e^{-i\Delta t}A+e^{i\Delta t}A^\dagger)\Big] \nonumber\\
&= -\frac{2\lambda^2}{\Delta}[A,A^\dagger],
\end{align}
so
\begin{equation}
\tfrac12[S',V] = -\frac{\lambda^2}{\Delta}[A,A^\dagger].
\end{equation}
Evaluating $[A,A^\dagger]$:
\begin{align}
A A^\dagger &= (a^\dagger b^\dagger\,|g\rangle\langle f|)(a b\,|f\rangle\langle g|)
= a^\dagger b^\dagger a b\;|g\rangle\langle g|
= a^\dagger a\;b^\dagger b\;|g\rangle\langle g|,\\[4pt]
A^\dagger A &= (a b\,|f\rangle\langle g|)(a^\dagger b^\dagger\,|g\rangle\langle f|)
= a b a^\dagger b^\dagger\;|f\rangle\langle f|
= a a^\dagger\;b b^\dagger\;|f\rangle\langle f|.
\end{align}
Thus
\begin{equation}
[A,A^\dagger] = a^\dagger a\,b^\dagger b\;|g\rangle\langle g| - a a^\dagger\,b b^\dagger\;|f\rangle\langle f|.
\end{equation}

Therefore
\begin{equation}
\tfrac12[S',V] = -\frac{\lambda^2}{\Delta}\Big(a^\dagger a\,b^\dagger b\;|g\rangle\langle g| - a a^\dagger\,b b^\dagger\;|f\rangle\langle f|\Big).
\end{equation}
Define
\begin{equation}
\chi \equiv \frac{\lambda^2}{\Delta},
\end{equation}
so
\begin{equation}
\tfrac12[S',V] = \chi\Big(a a^\dagger\,b b^\dagger\;|f\rangle\langle f| - a^\dagger a\,b^\dagger b\;|g\rangle\langle g|\Big).
\end{equation}

\bigskip
Adding this $\chi$-correction to the dispersive pieces yields the final effective Hamiltonian
\begin{equation}
\begin{aligned}
H_e &= -\frac{g_{A,c}^2}{\delta_a}\big(a^\dagger a\,|g\rangle\langle g| - a a^\dagger\,|e\rangle\langle e|\big)
- \frac{g_{B,c}^2}{\delta_b}\big(b^\dagger b\,|e\rangle\langle e| - b b^\dagger\,|f\rangle\langle f|\big)\\[4pt]
&\qquad\qquad + \chi\big(a a^\dagger\,b b^\dagger\,|f\rangle\langle f| - a^\dagger a\,b^\dagger b\,|g\rangle\langle g|\big),
\end{aligned}
\end{equation}
with $\displaystyle \chi=\frac{\lambda^2}{\Delta}$ and $\lambda=\dfrac{g_{A,c}g_{B,c}}{2}\!\left(\dfrac{1}{|\delta_a|}+\dfrac{1}{\delta_b}\right)$.

\bigskip
Thus the two-step SW eliminations (first eliminating $|e\rangle$, then eliminating the off-resonant $|g\rangle\leftrightarrow|f\rangle$ two-photon coupling) produce dispersive (AC-Stark) shifts $\propto g_{A,c}^2/\delta_a,\ g_{B,c}^2/\delta_b$ and the cross-Kerr interaction $\propto \chi\,n_a n_b$ (projected on the qutrit manifolds), with all algebraic steps shown above.

\section{Continuum-limit derivations}
\label{app:continuum}

\subsection{Derivation of the continuum limit of the kinetic (hopping) terms}
\label{app:cont_hop}

The kinetic part of our bipartite Bose--Hubbard model [Eq.~(\ref{eq:HEBH_again}) in the main text] is
\begin{equation}
    H_{\rm{hop}}=-\sum_{j}\left(t_{in}b^{\dagger}_{A,j}b_{B,j}+t_{out}b^{\dagger}_{B,j}b_{A,j+1}+\rm{h.c.}\right).
\end{equation}
The first term refers to hopping within an $A$--$B$ unit cell, and the second term to hopping from the $B$ site of the $j^{\rm th}$ unit cell to the $A$ site of the $(j+1)^{\rm th}$ unit cell. Again, it is a notable point that in our system, $t_{in}=t_{out}=g_l/4$.
We adopt the coordinate convention: at $x=0$ we place an $A$ site, at $x=a$ the corresponding $B$ site of the same unit cell (with $a$ the unit-cell length), and the next-cell $A$ site at $x=2a$.

We use the substitutions
\begin{equation}
b_{A,j}\approx\sqrt{\rho_0+\delta\rho_A(x)}e^{i\theta_A(x)},\qquad
b_{B,j}\approx\sqrt{\rho_0+\delta\rho_B(x)}e^{i\theta_B(x)},
\end{equation}
with $\delta\rho\simeq-\frac{1}{\pi}\partial_x\phi$. Define symmetric and antisymmetric fields
\begin{equation}
    \theta_{\pm}(x)=\frac{1}{\sqrt{2}}\left(\theta_A(x)\pm\theta_B(x)\right),\qquad
    \phi_{\pm}(x)=\frac{1}{\sqrt{2}}\left(\phi_A(x)\pm\phi_B(x)\right).
\end{equation}
We use Taylor expansions such as $\theta_B(x+a)\approx\theta_B(x)+a\partial_x\theta_B(x)+\frac{a^2}{2}\partial^2_{x}\theta_{B}+\cdots$.
Define
\begin{equation}
\Delta_{\rm{in}}(x)=\theta_A(x)-\theta_B(x+a),\qquad
\Delta_{\rm{out}}(x)=\theta_B(x+a)-\theta_A(x+2a),
\end{equation}
and $\theta_{A/B}$ in terms of $\theta_\pm$:
\begin{equation}
\theta_A(x)=\frac{\theta_+(x)+\theta_-(x)}{\sqrt{2}},\qquad
\theta_B(x)=\frac{\theta_+(x)-\theta_-(x)}{\sqrt{2}}.
\end{equation}

Upon substitution, the intra/inter contributions become
\begin{align}
    h_{\rm in}(x)&=-t_{\rm in}\left[b^{\dagger}(x)b(x+a)+\rm{h.c.}\right]\simeq-2t_{\rm in}\rho_0\cos\left(\Delta_{\rm in}(x)\right),\\
    h_{\rm out}(x)&=-t_{\rm out}\left[b^{\dagger}(x+a)b(x+2a)+\rm{h.c.}\right]\simeq-2t_{\rm out}\rho_0\cos\left(\Delta_{\rm out}(x)\right).
\end{align}
Expanding $\cos y\simeq1-\frac{y^2}{2}$ yields (up to a constant shift)
\begin{equation}
    t_{\rm in}\rho_0\left[\Delta\theta_{\rm in}\right]^2+t_{\rm out}\rho_0\left[\Delta\theta_{\rm out}\right]^2+O(a^3).
\end{equation}
Using the Taylor expansions inside $\Delta_{\rm in/out}$ gives
\begin{equation}
    \Delta_{\rm in}=\sqrt{2}\theta_-(x)-\frac{a}{\sqrt{2}}\left(\partial_x\theta_+(x)-\partial_x\theta_-(x)\right)+O(a^2),
\end{equation}
and
\begin{equation}
    \Delta_{\rm out}=-\sqrt{2}\theta_-(x)-\frac{a}{\sqrt{2}}\left(-\partial_x\theta_+(x)-3\partial_x\theta_-(x)\right)+O(a^2).
\end{equation}
Assuming symmetric hopping strengths $t_{\rm in}=t_{\rm out}=t$ (as in our system), converting $\sum_j\to\frac{1}{a}\int dx$, one obtains the continuum hopping Hamiltonian:
\begin{equation}
    H_{\rm hop}=\int dx\left[\frac{1}{2}M\theta_-^2+\frac{1}{2}s_+\left(\partial_x\theta_+\right)^2+\frac{1}{2}s_-\left(\partial_x\theta_-\right)^2+\lambda\left(\partial_x\theta_+\right)\left(\partial_x\theta_-\right)\right],
\end{equation}
where (for the symmetric-hopping case shown in the main text) $M=\frac{8t\rho_0}{a}$, $s_+=2t\rho_0a=\lambda$ and $s_-=10t\rho_0a$.

\subsection{The continuum limit derivation for the on-site interaction term(s)}
\label{app:cont_U}

In our bipartite Bose--Hubbard-like model, the on-site interaction term [Eq.~(\ref{eq:HEBH_again})] is
\begin{equation}
    H_U=\frac{U}{2}\sum_{j,\alpha=A,B}n_{j,\alpha}\left(n_{j,\alpha}-1\right).
\end{equation}
Using $n_{j,\alpha}=a\rho_{\alpha}(x_j)$ gives $n_{j,\alpha}(n_{j,\alpha}-1)=a^2\rho_\alpha^2-a\rho_\alpha$. Taking the continuum limit $\sum_j\to\frac{1}{a}\int dx$,
\begin{equation}
    H_U\supset\frac{Ua}{2}\int dx\left(\sum_{\alpha=A,B}\rho_\alpha(x)^2\right),
\end{equation}
where the linear term only gives global shifts/boundary terms and is dropped. Using the smooth part of the harmonic-fluid expansion $\rho_\alpha=\rho_0-\frac{1}{\pi}\partial_x\phi_\alpha+\cdots$, we obtain
\begin{equation}
    H_U=\frac{Ua}{2\pi^2}\int dx\left[\left(\partial_x\phi_+\right)^2+\left(\partial_x\phi_-\right)^2\right].
\end{equation}

\subsection{The continuum limit derivation for the chemical potential term(s) ($\mu_A\neq\mu_B$)}
\label{app:cont_mu}

The chemical potential term [Eq.~(\ref{eq:HEBH_again})] is
\begin{equation}
    H_{\mu}=-\sum_{j}\left(\mu_An_{j,A}+\mu_Bn_{j,B}\right).
\end{equation}
Since $n_{j,\alpha}\approx a\rho_\alpha(x_j)$,
\begin{equation}
    H_\mu=-\int dx\left(\mu_A\rho_A(x)+\mu_B\rho_B(x)\right).
\end{equation}
Define $\mu_0=(\mu_A+\mu_B)/2$ and $\Delta\mu=(\mu_A-\mu_B)/2$, so
\begin{equation}
    \mu_A\rho_A+\mu_B\rho_B=\mu_0(\rho_A+\rho_B)+\Delta\mu(\rho_A-\rho_B).
\end{equation}
Since the terms are linear in $\rho$, keep the smooth part and first harmonic
$\rho_\alpha=\rho_0-\frac{1}{\pi}\partial_x\phi_\alpha+A_1\cos(2\pi\rho_0 x-2\phi_\alpha)+\cdots$.
The uniform part produces only energy shifts and boundary terms and can be dropped.
The first harmonic yields
\begin{equation}
    H^{\rm osc}_{\mu}\sim\Delta\mu\int dx\left[\cos\!\left(\frac{\pi x}{a}\right)\,A_1\cos\!\left(2\pi\rho_0 x-2\phi_\alpha(x)\right)\right],
\end{equation}
where $\cos(\pi x/a)$ arises from staggered chemical potential ($(-1)^j$ at the lattice level).
This term can become nonoscillatory when
\begin{equation}
    \frac{\pi}{a}\pm2\pi\rho_0=0\ \Rightarrow\ \rho_0 a=\frac{n\mp1}{2},
\end{equation}
so at half-filling the system may support a commensurate CDW tendency.

\subsection{Continuum limit of the attractive cross--Kerr term}
\label{app:cont_chi}

\subsubsection*{Intra--cell cross--Kerr}
Starting from (\ref{eq:HEBH_again}) but in our $A$--$B$ unit cell:
\[
H_\chi^{(\text{intra})}=-\chi\sum_j n_{j,A}n_{j,B}
\;\approx\; -\chi\frac{1}{a}\int dx\;a^2\,\rho_A(x)\rho_B(x)
= -\chi a\int dx\;\rho_A(x)\rho_B(x).
\]
Inserting the expansion of the harmonic fluid and keep the smooth quadratic contribution:
\[
\rho_A\rho_B \supset \big(-\tfrac{1}{\pi}\partial_x\phi_A\big)\big(-\tfrac{1}{\pi}\partial_x\phi_B\big)
= \frac{1}{\pi^2}(\partial_x\phi_A)(\partial_x\phi_B).
\]
Hence the intra--cell smooth piece is
\begin{equation}\label{eq:intra_smooth}
H_{\chi}^{(\text{intra}),\text{cont}}
= -\chi a\int dx\;\frac{1}{\pi^2}(\partial_x\phi_A)(\partial_x\phi_B).
\end{equation}
Using \(\partial_x\phi_A=(\partial_x\phi_+ + \partial_x\phi_-)/\sqrt2\) and
\(\partial_x\phi_B=(\partial_x\phi_+ - \partial_x\phi_-)/\sqrt2\) we obtain
\[
(\partial_x\phi_A)(\partial_x\phi_B)=\tfrac12\big[(\partial_x\phi_+)^2 - (\partial_x\phi_-)^2\big],
\]
so that
\begin{equation}\label{eq:intra_cont_final}
H_{\chi}^{(\text{intra}),\text{cont}}
= -\frac{\chi a}{2\pi^2}\int dx\;(\partial_x\phi_+)^2 \;+\; \frac{\chi a}{2\pi^2}\int dx\;(\partial_x\phi_-)^2 .
\end{equation}

The oscillatory term arising from the product of the first harmonics in the intra-cell case produces
\[
A_1^2\cos(2\pi\rho_0 x - 2\phi_A)\cos(2\pi\rho_0 x - 2\phi_B)
= \frac{A_1^2}{2}\Big[\cos\big(2(\phi_A-\phi_B)\big) + \cos\big(4\pi\rho_0 x - 2(\phi_A+\phi_B)\big)\Big].
\]
The first term is non--oscillatory (no explicit $x$-dependence) and equals $\cos\big(2(\phi_A-\phi_B)\big)=\cos(2\sqrt2\,\phi_-)$. Thus intra--cell $\chi$ generates a local antisymmetric cosine:
\begin{equation}\label{eq:intra_cosine}
H_{\chi}^{(\text{intra}),\text{osc}}
\supset -g_{\chi}^{(\text{intra})}\int dx\;\cos\!\big(2\sqrt2\,\phi_-(x)\big),\qquad
g_{\chi}^{(\text{intra})}\propto \chi\,a\,\frac{A_1^2}{2}.
\end{equation}

\subsubsection*{Inter--cell cross--Kerr: continuum limit}
Starting from (\ref{eq:HEBH_again})
\[
H_\chi^{(\text{inter})}=-\chi\sum_j n_{j,B}n_{j+1,A}
\;\approx\; -\chi\frac{1}{a}\int dx\;a^2\,\rho_B(x)\rho_A(x+a)
= -\chi a\int dx\;\rho_B(x)\rho_A(x+a).
\]
Expanding \(\rho_A(x+a)=\rho_A(x)+a\partial_x\rho_A(x)+\cdots\). to leading order in gradients, the smooth product again yields the same quadratic gradient term as the intra--cell contribution:
\[
\rho_B(x)\rho_A(x+a)\supset \frac{1}{\pi^2}(\partial_x\phi_B)(\partial_x\phi_A) + O(a\partial^3\phi),
\]
and therefore
\begin{equation}\label{eq:inter_cont_final}
H_{\chi}^{(\text{inter}),\text{cont}}
= -\frac{\chi a}{2\pi^2}\int dx\;(\partial_x\phi_+)^2 \;+\; \frac{\chi a}{2\pi^2}\int dx\;(\partial_x\phi_-)^2 \;+\; O(a^2\partial^3\phi).
\end{equation}

The product of first harmonics with the spatial shift $a$ contains terms analogous to the intra--cell case. Expanding $\phi_A(x+a)\approx\phi_A(x)$ to leading order yields the same local antisymmetric cosine $\cos(2\sqrt2\,\phi_-)$, with an amplitude of the same order as the intra--cell contribution, up to microscopic phase factors.

\subsubsection*{Combined result for $\chi_{\rm intra}=\chi_{\rm inter}=\chi$}
Adding \eqref{eq:intra_cont_final} and \eqref{eq:inter_cont_final}, or simply doubling the intra--cell contribution, gives the leading quadratic continuum cross--Kerr Hamiltonian:
\begin{equation}
H_{\chi}^{\mathrm{cont}}
= -\frac{\chi_\Sigma a}{2\pi^2}\int dx\;(\partial_x\phi_+)^2 \;+\; \frac{\chi_\Sigma a}{2\pi^2}\int dx\;(\partial_x\phi_-)^2
= -\frac{\chi a}{\pi^2}\int dx\;(\partial_x\phi_+)^2 \;+\; \frac{\chi a}{\pi^2}\int dx\;(\partial_x\phi_-)^2,
\end{equation}
where $\chi_\Sigma=\chi_{\rm intra}+\chi_{\rm inter}=2\chi$. Thus the cross--Kerr shifts the compressibility coefficients with opposite signs in the $\pm$ channels.

The combined lowest non--oscillatory cosine from intra+inter is
\begin{equation}
H_{\chi}^{\mathrm{osc}} \;\supset\; -\,g_{\chi,-}\int dx\;\cos\!\big(2\sqrt2\,\phi_-(x)\big),
\qquad g_{\chi,-} \sim \tfrac{1}{2}\chi a A_1^2,
\end{equation}
where $A_1$ is the (nonuniversal) amplitude of the first density harmonic in \eqref{eq:rho_harmonic} and the exact numerical prefactor may depend on microscopic overlap integrals.

\subsubsection*{Compressibility shifts (combined with on--site \texorpdfstring{U}{U})}
If the on--site interaction contributes $c_U$ to each channel (with $c_U=Ua/(2\pi^2)$ in the usual mapping), then including the cross--Kerr yields
\[
\,c_\pm \;=\; c_U \;\mp\; \frac{\chi_\Sigma a}{2\pi^2}
\;=\; \frac{a}{2\pi^2}\big(U \mp \chi_\Sigma\big)
\;=\; \frac{a}{2\pi^2}\big(U \mp 2\chi\big)\,.
\]
Equivalently, in the equal-$\chi$ case:
\[
\,c_+ = \frac{a}{2\pi^2}(U-2\chi),\qquad c_- = \frac{a}{2\pi^2}(U+2\chi).\,
\]

\subsubsection*{Explicit \texorpdfstring{$\chi$}{\ensuremath{\chi}}-dependence of \texorpdfstring{$K_+$}{K_+} (symmetric-hopping)}
For symmetric hopping $s_+=s$ and the total-mode Luttinger parameter becomes
\begin{align}
K_+(\chi)
&= \sqrt{\frac{s}{c_+}}
= \sqrt{\frac{s}{\dfrac{a}{2\pi^2}(U-2\chi)}} \nonumber\\[6pt]
&= \sqrt{\frac{2\pi^2 s}{a\,(U-2\chi)}} \label{eq:Kplus_generic}\\[6pt]
&=\sqrt{\frac{2\pi^2 (2\pi a\rho_0 t)}{a\,(U-2\chi)}}
= \sqrt{\frac{4\pi^3 \rho_0 t}{\,U-2\chi\,}} \nonumber\\[6pt]
&= 2\pi^{3/2}\,\sqrt{\frac{\rho_0 t}{\,U-2\chi\,}}. \label{eq:Kplus_t}
\end{align}
Thus (for $U-2\chi>0$) $K_+$ grows monotonically with attractive $\chi>0$.

\section{Effective single-component Luttinger Hamiltonian for the \texorpdfstring{$+$}{+} mode}
\label{app:derive_LL_plus}

We start from the continuum Hamiltonian used in the main text (long-wavelength limit)
\[
\begin{split}
H_{\mathrm{hop}} &= \int dx\ \bigg\{
\frac{1}{2}M\,\theta_-^2
+\frac{1}{2}s_+(\partial_x\theta_+)^2
+\frac{1}{2}s_-(\partial_x\theta_-)^2
+\lambda(\partial_x\theta_+)(\partial_x\theta_-)
\bigg\} \\
&\qquad\qquad + \int dx\ \bigg\{
\frac{\kappa_+}{2}(\partial_x\phi_+)^2
+\frac{\kappa_-}{2}(\partial_x\phi_-)^2
\bigg\},
\end{split}
\]
with
\[
\kappa_\pm \equiv \frac{Ua\mp 2\chi}{2\pi^2},
\]
and the canonical commutator
\[
[\phi_\alpha(x),\,\partial_y\theta_\beta(y)] = i\pi\,\delta_{\alpha\beta}\,\delta(x-y),
\qquad \alpha,\beta\in\{+,-\}.
\]

We use Euclidean time \(\tau\) and Fourier transform fields as
\[
\varphi(x,\tau)=\int\frac{dk\,d\omega}{(2\pi)^2}\,e^{i(kx-\omega\tau)}\,\varphi(k,\omega),
\]
with the reality condition \(\varphi(-k,-\omega)=\varphi(k,\omega)^*\).
For compactness denote \(\varphi(+)\equiv\varphi(k,\omega)\) and \(\varphi(-)\equiv\varphi(-k,-\omega)\).

From the canonical structure the Euclidean kinetic term is
\[
S_{\mathrm{kin}}=\int\frac{d\omega\,dk}{(2\pi)^2}\;\frac{i k \omega}{\pi}\Big[\phi_+(-)\theta_+(+)+\phi_-(-)\theta_- (+)\Big].
\]
The potential (gradient and mass) piece is
\[
\begin{aligned}
S_{\mathrm{pot}}=\tfrac12\int\frac{d\omega\,dk}{(2\pi)^2}\big[ &
\kappa_+ k^2\,\phi_+(-)\phi_+(+)+\kappa_- k^2\,\phi_-(-)\phi_-(+) \\
&+ s_+ k^2\,\theta_+(-)\theta_+(+) + (s_- k^2 + M)\,\theta_-(-)\theta_-(+) \\
&+ \lambda k^2\big(\theta_+(-)\theta_-(+)+\theta_-(-)\theta_+(+)\big)
\big].
\end{aligned}
\]
The full quadratic action is \(S=S_{\mathrm{kin}}+S_{\mathrm{pot}}\).

Group the antisymmetric variables into the 2-vector \(x\) and symmetric into \(y\):
\[
x\equiv\begin{pmatrix}\phi_-\\[4pt]\theta_-\end{pmatrix},\qquad
y\equiv\begin{pmatrix}\phi_+\\[4pt]\theta_+\end{pmatrix}.
\]
For each \((k,\omega)\) the quadratic integrand can be written as
\[
\tfrac12\,x^T A x + x^T B y + \tfrac12\,y^T C y,
\]
with \(2\times2\) matrices \(A,B,C\). Define
\[
a\equiv \kappa_- k^2,\qquad b\equiv \frac{i k \omega}{\pi},\qquad c\equiv s_- k^2 + M.
\]
Then
\[
A=\begin{pmatrix} a & b \\[4pt] b & c \end{pmatrix}
= \begin{pmatrix}\kappa_- k^2 & \dfrac{i k \omega}{\pi} \\[6pt] \dfrac{i k \omega}{\pi} & s_- k^2 + M \end{pmatrix},
\qquad
B=\begin{pmatrix} 0 & 0 \\[4pt] 0 & \lambda k^2 \end{pmatrix},
\qquad
C=\begin{pmatrix} \kappa_+ k^2 & \dfrac{i k \omega}{\pi} \\[6pt] \dfrac{i k \omega}{\pi} & s_+ k^2 \end{pmatrix}.
\]

Integrating out \(x\) yields the exact effective action for \(y\):
\[
S_{\mathrm{eff}}(y)=\tfrac12\,y^T\big(C - B^T A^{-1} B\big)y.
\]
For \(A=\begin{pmatrix} a & b \\ b & c \end{pmatrix}\) the inverse is
\[
A^{-1}=\frac{1}{\Delta}\begin{pmatrix} c & -b \\[4pt] -b & a \end{pmatrix},\qquad
\Delta\equiv ac-b^2.
\]
Compute
\[
b^2=\Big(\frac{i k \omega}{\pi}\Big)^2=-\frac{(k\omega)^2}{\pi^2},
\]
and
\[
\Delta = a c - b^2 = \kappa_- k^2 (s_- k^2 + M) + \frac{(k\omega)^2}{\pi^2}.
\]
The \((2,2)\) element of \(A^{-1}\) is
\[
(A^{-1})_{22}=\frac{a}{\Delta}
=\frac{\kappa_- k^2}{\kappa_- k^2 (s_- k^2 + M) + \dfrac{(k\omega)^2}{\pi^2}}.
\]
Since \(B\) has only the (2,2) entry nonzero,
\[
B^T A^{-1} B = \begin{pmatrix} 0 & 0 \\[4pt] 0 & (\lambda k^2)^2 (A^{-1})_{22} \end{pmatrix}.
\]

The frequency-/momentum-dependent self-energy entering the \(\theta_+\)-kernel is
\[
\Delta_{\theta_+}(k,\omega) = (\lambda k^2)^2 (A^{-1})_{22}
= \frac{\lambda^2 k^4\,(\kappa_- k^2)}
{\kappa_- k^2 (s_- k^2 + M) + \dfrac{(k\omega)^2}{\pi^2}}.
\]
Dividing numerator and denominator by \(\kappa_- k^2\) (for \(k\neq0\)) gives the simpler algebraic form
\[
\Delta_{\theta_+}(k,\omega) = \frac{\lambda^2 k^4}{s_- k^2 + M + \dfrac{\omega^2}{\pi^2\kappa_-}}.
\]
Therefore the effective \(+\)-block is
\[
C_{\mathrm{eff}}(k,\omega)=
\begin{pmatrix}
\kappa_+ k^2 & \dfrac{i k \omega}{\pi} \\[6pt]
\dfrac{i k \omega}{\pi} & s_+ k^2 - \Delta_{\theta_+}(k,\omega)
\end{pmatrix}.
\]

In the infrared limit \(k\to0,\ \omega\to0\) with \(\omega\sim v k\), and assuming the antisymmetric phase is pinned by a large mass \(M\) (so \(M\) dominates the denominator), we have
\[
s_- k^2 + M + \frac{\omega^2}{\pi^2\kappa_-} \approx M,
\]
and hence
\[
\Delta_{\theta_+}(k,\omega)\simeq \frac{\lambda^2}{M}\,k^4.
\]
Thus the \(\theta_+\) kernel becomes
\[
 \simeq s_+ k^2 - \frac{\lambda^2}{M}\,k^4,
\]
which in real space corresponds to the higher-derivative operator
\[
\delta H = \frac{1}{2}\,\frac{\lambda^2}{M}\int dx\;(\partial_x^2\theta_+)^2.
\]
This operator is irrelevant at long wavelengths compared with \((\partial_x\theta_+)^2\).

Matching the leading coefficients to the standard single-component Luttinger Hamiltonian
\[
H_{\mathrm{eff},+}=\frac{u_+}{2\pi}\int dx\ \left[ K_+(\partial_x\theta_+)^2 + \frac{1}{K_+}(\partial_x\phi_+)^2 \right],
\]
we obtain to leading order
\[
u_+\simeq\sqrt{s_+\,\kappa_+}, \qquad K_+\simeq\sqrt{\frac{s_+}{\kappa_+}}.
\]
Corrections to these expressions are suppressed by powers of \(1/M\) and by additional factors of \(k\); therefore, in the regime where \(\theta_-\) is strongly pinned, the low-energy physics is described by the above single-component Luttinger Hamiltonian for the \(+\) mode.

\section{Correlation functions from the single-component Luttinger liquid}
\label{app:correlators}

Here we provide minimal derivations of the long-distance forms of $g^{(1)}(x)$, $g^{(2)}(x)$, and $C_{zz}(x)$ used in the main text. We assume the antisymmetric sector is gapped, so only the $+$ mode remains gapless at long distances.

\subsection{First-order coherence $g^{(1)}(x)$}
In the LL regime, the low-energy bosonic field has the form
\begin{equation}
\Psi(x)\sim \sqrt{\rho_0}\,e^{i\theta_+(x)}\times(\text{short-distance factors}),
\end{equation}
so
\begin{equation}
g^{(1)}(x)=\frac{\langle \Psi^\dagger(x)\Psi(0)\rangle}{\rho_0}\sim \left\langle e^{i(\theta_+(x)-\theta_+(0))}\right\rangle.
\end{equation}
For a Gaussian LL action, $\langle e^{iA}\rangle=\exp[-\frac12\langle A^2\rangle]$, hence
\begin{equation}
g^{(1)}(x)\sim \exp\!\left[-\frac12\langle(\theta_+(x)-\theta_+(0))^2\rangle\right]
\sim \left(\frac{a}{|x|}\right)^{\frac{1}{4K_+}}.
\label{eq:app_g1_final}
\end{equation}

\subsection{Spatial density-density correlation \texorpdfstring{$g^{(2)}(x)$}{g^{(2)}(x)}}
Writing $n(x)=\rho_0+\delta n(x)$, one has
\begin{equation}
g^{(2)}(x)=\frac{\langle n(x)n(0)\rangle}{\rho_0^2}
=1+\frac{\langle \delta n(x)\delta n(0)\rangle}{\rho_0^2}.
\end{equation}
In LL theory,
\begin{equation}
\delta n(x)\simeq -\frac{1}{\pi}\partial_x\phi_+(x)+A_{2k}\cos\!\big(2\pi\rho_0 x-2\phi_+(x)\big)+\cdots.
\end{equation}
The smooth part gives the universal tail $\langle \partial_x\phi_+(x)\partial_x\phi_+(0)\rangle\sim K_+/(2x^2)$, yielding
\begin{equation}
\langle \delta n(x)\delta n(0)\rangle
\simeq
-\frac{K_+}{2\pi^2}\frac{1}{x^2}
+
A_{2k}\frac{\cos(2\pi\rho_0 x)}{|x|^{2K_+}}+\cdots,
\end{equation}
and therefore
\begin{equation}
g^{(2)}(x)\simeq
1-\frac{K_+}{2\pi^2\rho_0^2}\frac{1}{x^2}
+\frac{A_{2k}}{\rho_0^2}\frac{\cos(2\pi\rho_0 x)}{|x|^{2K_+}}+\cdots.
\label{eq:app_g2_final}
\end{equation}

\subsection{Qubit magnetization correlator \texorpdfstring{$C_{zz}(x)$}{C_{zz}(x)}}
After projecting to the low-energy manifold and integrating out the gapped $-$ sector, any local operator such as $\sigma^z(x)$ can be expanded in symmetry-allowed operators of the remaining gapless LL:
\begin{equation}
\sigma^z(x)=c_0+c_1\,\partial_x\phi_+(x)
+c_2\cos\!\big(2\pi\rho_0 x-2\phi_+(x)\big)+\cdots,
\label{eq:app_sigmaz_expand}
\end{equation}
where the coefficients $c_i$ are nonuniversal and depend on microscopic dressing (including short-ranged $-$-sector fluctuations).
Using the same LL correlators as above gives the connected correlator
\begin{equation}
C_{zz}(x)\equiv \langle \sigma^z(x)\sigma^z(0)\rangle-\langle\sigma^z\rangle^2
\simeq
-\frac{A(\chi)}{x^2}+B(\chi)\frac{\cos(2\pi\rho_0 x)}{|x|^{2K_+}}+\cdots.
\label{eq:app_Czz_final}
\end{equation}

\section{Appendix figures: diagnostics and interpretation}
\label{app:appendix_figs}

The following figures provide diagnostic checks and supplementary views of the correlation functions discussed in the main text.

\subsection{The universal $1/x^2$ in $C_{zz}(x)$}
At long distances, the connected correlator has the universal LL smooth tail $C_{zz}(x)\sim -A(\chi)/x^2$ (up to oscillatory corrections). Multiplying by $(x/a)^2$ isolates the prefactor and makes the asymptotic regime visible on linear axes: $(x/a)^2 C_{zz}(x)\to -A(\chi)$ as $x/a\to\infty$. Deviations at small $x/a$ reflect short-range(UV) physics beyond the Luttinger-Liquid regime and the residual $2k_F$ oscillatory part.

\begin{figure}[t]
  \centering
  \includegraphics[width=0.9\linewidth]{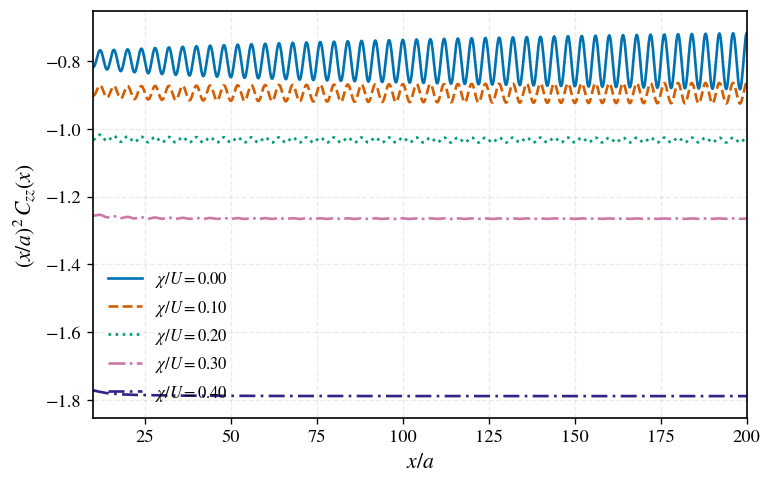}
  \caption{Diagnostic plot of $(x/a)^2 C_{zz}(x)$ emphasizing the approach to a constant plateau at large $x/a$, which encodes the prefactor of the universal $1/x^2$ smooth component of the Luttinger liquid correlator. The residual oscillations arise from the subleading $2k_F$ term and finite short-distance cutoff effects.}
  \label{fig:app_x2Czz}
\end{figure}

\subsection{Photon-qubit equivalence in the LL regime}
In the LL regime with a single gapless $+$ mode, both photonic and qubit operators couple to the same long-wavelength phase field, so their normalized first-order coherences share the same algebraic exponent. Plotting the normalized correlators on the same axes provides a direct internal consistency check of the projection and of the single-mode assumption.

\begin{figure}[t]
  \includegraphics[width=0.9\linewidth]{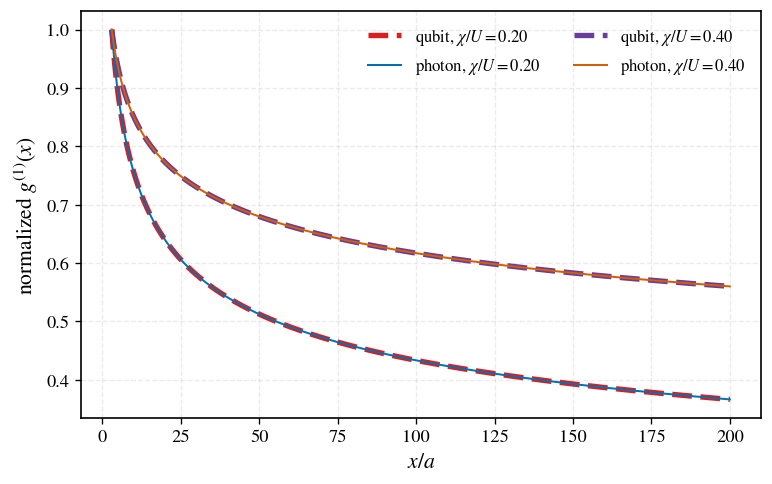}
  \caption{Normalized first-order coherence in the photon and qubit channels. The collapse of the long-distance scaling confirms that both channels probe the same gapless $+$ mode, consistent with the effective single-component Luttinger description used in the main text.}
  \label{fig:app_channel_equiv}
\end{figure}

\bibliographystyle{apsrev4-2}   
\bibliography{references3}

\end{document}